\documentclass[letterpaper,twocolumn,10pt]{article}
\usepackage{usenix2024_SOUPS}

\usepackage{tikz}
\usepackage{amsmath}
\usepackage{comment}
\usepackage{subcaption}

\begin{document}

\date{}

\title{\Large \bf The PIONEER Project: A PrIvacy companion \\ for mOtivatioN and knowlEdge transfER}

\def\plainauthor{Simon Althaus, Nina Gerber, Sara Hahn, Andreas Heinemann, Angela Menig, Max M\"uhlh\"auser, Christian Reuter, Ephraim Zimmer}

\author{
{\rm 
Simon Althaus$^\dagger$$^\ddagger$,
Nina Gerber$^\dagger$$^\ddagger$
Sara Hahn$^\star$$^\dagger$$^\ddagger$, 
Andreas Heinemann$^\star$$^\ddagger$, 
} \\
{\rm Angela Menig$^\dagger$$^\ddagger$,
Max M\"uhlh\"auser$^\dagger$$^\ddagger$,
Christian Reuter$^\dagger$$^\ddagger$,
Ephraim Zimmer$^\dagger$$^\ddagger$}\\
$^\dagger$ Technical University of Darmstadt \\
$^\star$University of Applied Sciences Darmstadt \\
$^\ddagger$ATHENE – National Research Center for Applied Cybersecurity
} %

\maketitle
\thecopyright

\begin{abstract}
Remaining control over their private data is one of the key challenges in this century for users 
We know from prior work that users are often neither in a position to fully grasp the content of the usually complicated texts, nor are they motivated to spend the time necessary to do so.
We report on the progress made by the PIONEER project on a privacy support tool that combines knowledge transfer and persuasive elements to increase users' privacy awareness and motivation; thus empowering them to more privacy sovereignty. Throughout the research and design process, we consider user group specifics that may result in different requirements, e.g., for children, adolescents, parents, or elderly people. We further target sustainable behavior change by addressing different states of change, precisely: spark initial motivation, facilitate the creation of new habits, and encourage habituation of these habits in the long term (volition). Finally, we provide a privacy support tool demonstrator that can be utilized for research and education purposes, e.g., in school contexts.
\end{abstract}

\section{Introduction}

Remaining control over their private data is one of the key challenges in this century for users \cite{Pew2019Americans,acquisti2020secrets,story2021awareness}. Attempts by legislators to help users meet this challenge such as the General Data Protection Regulation (GDPR) \cite{european_commission_regulation_2016} in the EU are often based on the concept of “informed consent”. This assumes that users can decide according to their actual interest if they are provided with sufficient information, e.g., about how and by which actors their data will be processed. In practice, this is reflected in extensive privacy statements or pop-up windows, e.g., informing about the use of cookies on various websites.

We know from prior work that users are often neither in a position to fully grasp the content of the usually complicated texts, nor are they motivated to spend the time necessary to do so \cite{habib2020,Gray2021,obarBiggestLieInternet2020,Utz2019,Machuletz2020,Nouwens2020,Grassl2021}. Hence, a new concept is required to ensure that users can make truly sovereign decisions about the handling of their data in everyday digital life, which (1) conveys the necessary knowledge, (2) strengthens the users’ motivation to acquire competencies in this area and apply these in everyday digital life, especially taking into account (3) specifics that evolve from different user groups, like elderly people or adolescents.

We are thus striving for the user-centered as well as user group diacritic development of a digital privacy companion in the form of a mobile app that combines all three elements. 
\begin{enumerate}
    \item The app provides the necessary knowledge about digital contexts to enable users to actually make self-determined, informed decisions about how to handle their data.
    \item We further apply psychological motivation theories to address psychological needs (e.g., autonomy, competence, relatedness) that drive human behavior but have sparsely been considered in the Usable Security and Privacy context so far. In addition, we build on the Persuasive System Design framework \cite{OinasKukkonen2009} to identify persuasive principles that support users to show and maintain privacy-aware behavior in the long term.
    \item Our user-centered development explicitly examines and considers specifics of the aforementioned goals and methods that derive from different user groups, since, e.g., elderly people are driven by different psychological needs than younger generations, and children and adolescents need to be addressed with a different level of knowledge about digital contexts than adults.
\end{enumerate}

We provide an overview of the research conducted by the PIONEER project in the remainder.

\section{Current Investigations \& Results}

\begin{figure}
    \centering
    \begin{subfigure}[b]{0.4\columnwidth}
        \centering
        \includegraphics[width=\textwidth]{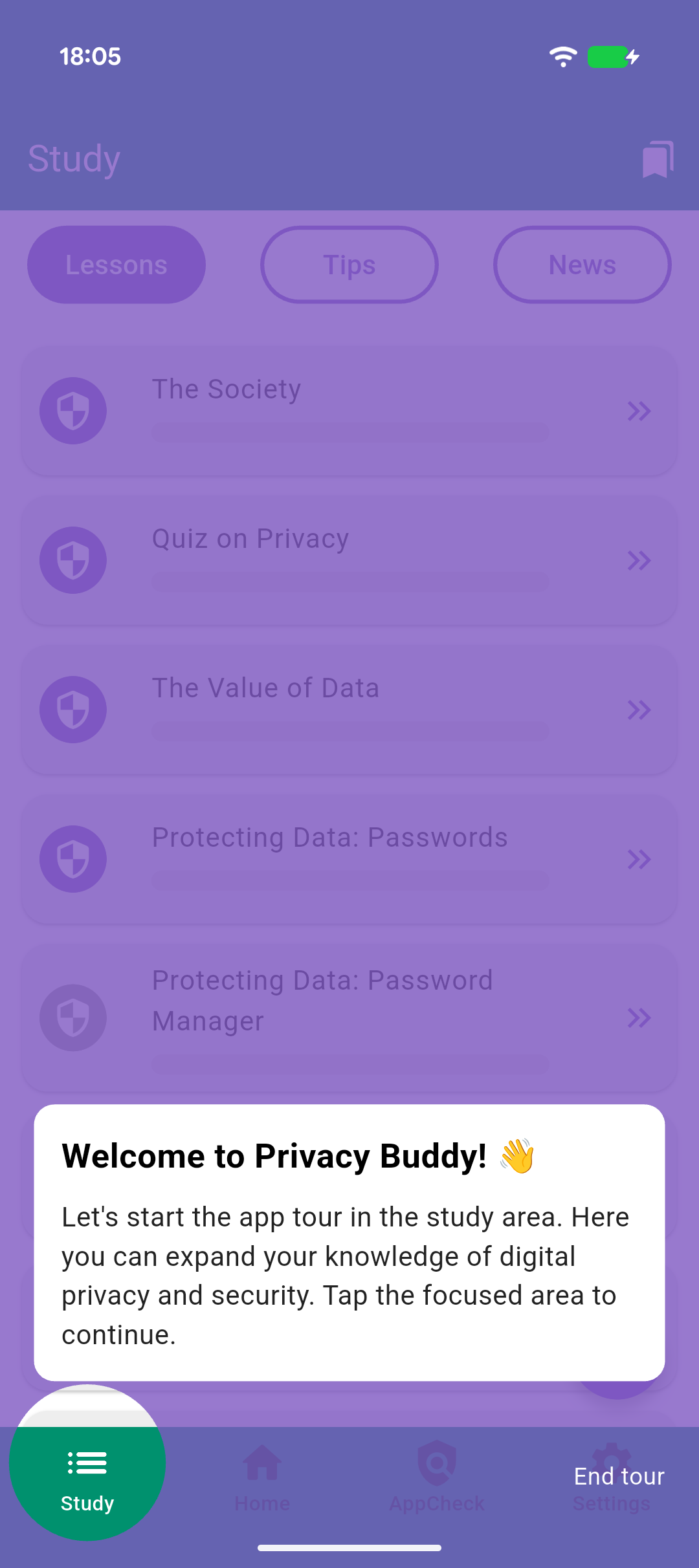}
    \end{subfigure}
    \hfill
    \begin{subfigure}[b]{0.4\columnwidth}
        \centering
        \includegraphics[width=\textwidth]{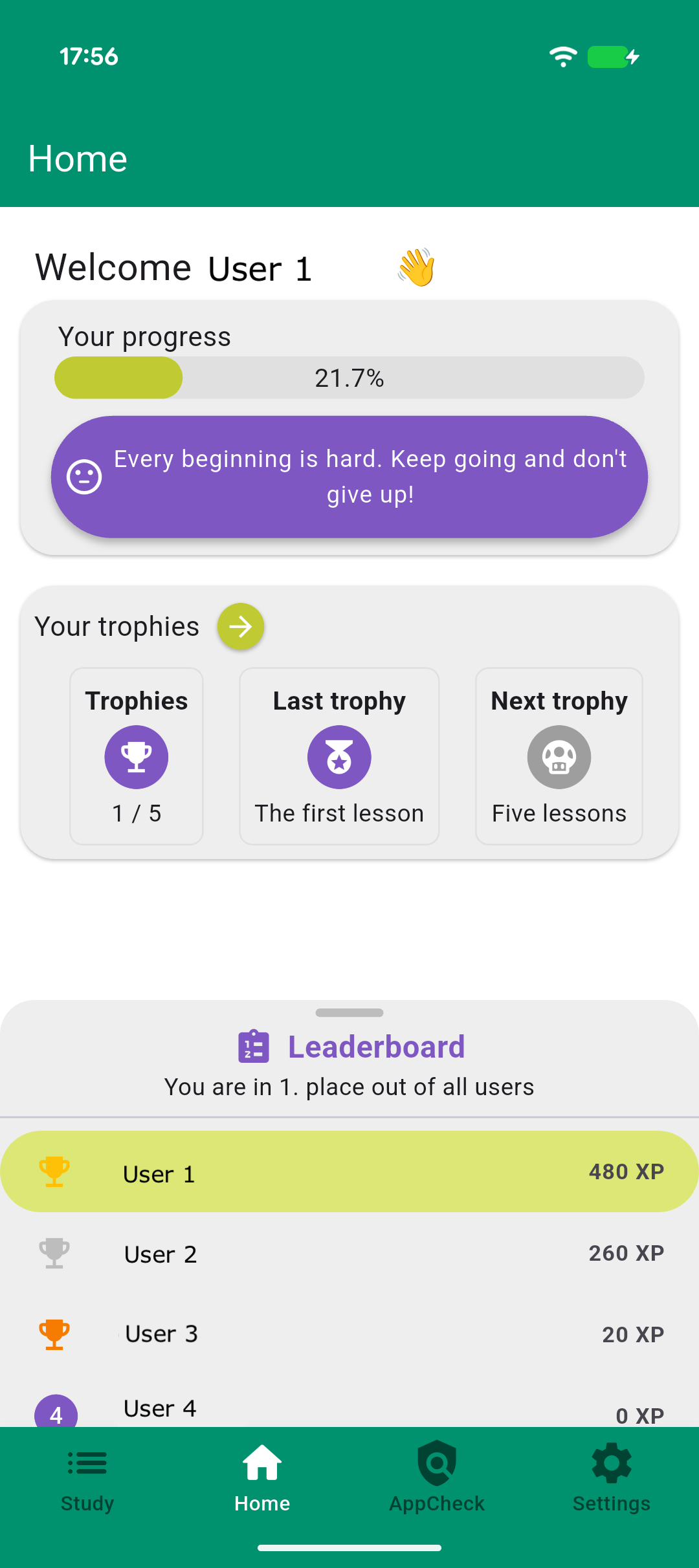}
    \end{subfigure}
    \caption{Screenshots of our privacy companion.}
    \label{fig:screenshots}
\end{figure}

\subsection{Technical Realization} %

We developed an initial prototype companion app for Android. 
We collected and reviewed relevant literature on the implementation of persuasive mechanisms as well as on the customization of mobile applications with respect to different user groups.
Building on several iterations of usability evaluations, the prototype was improved through redesign efforts and the introduction of new features like multilingual support, cross-platform availability, survey integration, and a guided on-boarding experience.
In addition, pipelines for machine learning-based features enable the analysis of privacy policies, chat bot support, and assistance for creation of teaching content.
We further implemented a modular and extensible framework for mobile app analysis, which serves analysis results to the companion app about the privacy of mobile Android apps.
In addition, research on Android’s Data Safety Section revealed fewer discrepancies than previously reported when applying refined definitions. Complementing this, new user analysis and Android tracing frameworks were designed and implemented \cite{slicedroid,wootdroid}. %

\subsection{Persuasive Systems Design and User Groups} %

We conducted multiple studies to iteratively refine the companion app within a user-centered design process. These included (1) a design thinking study to analyze and derive persuasive app features based on psychological needs, (2) a controlled online experiment investigating visualizations of privacy policies, and (3) expert focus groups to identify relevant privacy situations and data flows based on contextual integrity.
Our key findings from the focus groups are that structuring privacy information, ensuring verifiability, and highlighting the credibility of providers is prioritized over gamification elements which seem unsuited in the privacy context.
The online experiment showed that visual metaphors displaying which data flows towards third parties and which is stored only by the data collector facilitate understanding and are preferred by users.
In addition, we focused on children and teenagers as specific user groups and conducted interviews to understand their privacy support needs and requirements. Based on these findings, we created a video directly addressing children and teens to support them in managing their privacy on social media. We also added lessons to the prototype based on privacy training materials validated in a previously conducted user survey study. Furthermore, we conducted a systematic literature review to identify the most promising persuasive mechanisms and their combinations, identifying primary task support as the most promising, followed by dialogue and social features.

\subsection{Psychological Needs} %

Our ongoing research entails several empirical and design-oriented activities.
(1) We conducted an interview study with young adults to examine their privacy-related decision-making and to investigate how these decisions relate to underlying psychological needs. We found that their key motivation, wanting to feel competent or entertained, or maintain close contact with friends, outweighed their privacy concerns.
(2) We further performed usability evaluations of the current version of our app in order to assess user interaction and identify opportunities for improvement.
(3) In addition, we held focus groups with both users and domain experts to explore perspectives on privacy decision-making in greater depth and confirmed the information type to be of great importance. We created a scenario set of daily privacy scenarios with varying types that will be used in a follow-up quantitative study. %
(4) We conducted co-creation workshops aimed at developing need-sensitive design approaches for trustworthy permission approval processes in smartphone analysis. The outcomes of these workshops inform iterative design improvements and were evaluated in a quantitative follow-up study investigating the relationship of need fulfillment and trust in our app. Such trust-building elements include an overview of apps accessing personal data, permission management options, a privacy score, and privacy certificates.

\section*{Acknowledgments}
This research work was supported by the National Research Center for Applied Cybersecurity ATHENE. ATHENE is funded jointly by the German Federal Ministry of Research, Technology and Space and the Hessian Ministry of Science and Research, Arts and Culture.

\bibliographystyle{plain}
\bibliography{usenix2024_SOUPS}

@misc{slicedroid,
  doi = {10.5281/zenodo.15784276},
  url = {https://zenodo.org/doi/10.5281/zenodo.15784276},
  author = {Alexopoulos, Nikolaos and Althaus, Simon and Spinellis, Diomidis},
  title = {SliceDroid: Towards Reconstructing Android Application I/O Behaviors from Kernel Traces},
  publisher = {Zenodo},
  year = {2025},
  copyright = {Creative Commons Attribution 4.0 International}
}

@misc{wootdroid,
  doi = {10.48550/ARXIV.2604.27830},
  url = {https://arxiv.org/abs/2604.27830},
  author = {Althaus, Simon and Alexopoulos, Nikolaos and Mühlhäuser, Max and Reuter, Christian and Zimmer, Ephraim},
  title = {WOOTdroid: Whole-system Online On-device Tracing for Android},
  publisher = {arXiv},
  year = {2026},
  copyright = {arXiv.org perpetual, non-exclusive license}
}

@misc{Pew2019Americans,
     title = {Americans and Privacy: Concerned, Confused and Feeling Lack of Control Over Their Personal Information},
     author = {Brooke Auxier and Lee Rainie and Monica Anderson and Andrew Perrin and Madhu Kumar and Erica Turner},
    publisher = {Pew Research Center},
    date= {2019-11},
    url={https://www.pewresearch.org/internet/wp-content/uploads/sites/9/2019/11/Pew-Research-Center_PI_2019.11.15_Privacy_FINAL.pdf},
    urldate={2023-09-01},
year = {2019}
}

@article{acquisti2020secrets,
  title={Secrets and likes: the drive for privacy and the difficulty of achieving it in the digital age},
  author={Acquisti, Alessandro and Brandimarte, Laura and Loewenstein, George},
  journal={Journal of Consumer Psychology},
  volume={30},
  number={4},
  pages={736--758},
  year={2020},
  publisher={Wiley Online Library}
}

@article{story2021awareness,
  title={Awareness, adoption, and misconceptions of web privacy tools},
  author={Story, Peter and Smullen, Daniel and Yao, Yaxing and Acquisti, Alessandro and Cranor, Lorrie Faith and Sadeh, Norman and Schaub, Florian},
  journal={Proceedings on Privacy Enhancing Technologies},
  volume={2021},
  number={3},
  pages={308--333},
  year={2021}
}

@inproceedings{habib2020,
author = {Habib, Hana and Pearman, Sarah and Wang, Jiamin and Zou, Yixin and Acquisti, Alessandro and Cranor, Lorrie Faith and Sadeh, Norman and Schaub, Florian},
title = {"It's a Scavenger Hunt": Usability of Websites' Opt-Out and Data Deletion Choices},
year = {2020},
isbn = {9781450367080},
publisher = {Association for Computing Machinery},
address = {New York, NY, USA},
url = {https://doi.org/10.1145/3313831.3376511},
doi = {10.1145/3313831.3376511},
booktitle = {Proceedings of the 2020 CHI Conference on Human Factors in Computing Systems},
pages = {1–12},
numpages = {12},
location = {Honolulu, HI, USA},
series = {CHI '20}
}

@inproceedings{Gray2021,
author = {Gray, Colin M. and Santos, Cristiana and Bielova, Nataliia and Toth, Michael and Clifford, Damian},
title = {Dark Patterns and the Legal Requirements of Consent Banners: An Interaction Criticism Perspective},
year = {2021},
isbn = {9781450380966},
publisher = {Association for Computing Machinery},
address = {New York, NY, USA},
url = {https://doi.org/10.1145/3411764.3445779},
doi = {10.1145/3411764.3445779},
booktitle = {Proceedings of the 2021 CHI Conference on Human Factors in Computing Systems},
articleno = {172},
numpages = {18},
location = {Yokohama, Japan},
series = {CHI '21}
}

@article{obarBiggestLieInternet2020,
     title = {The biggest lie on the {Internet}: ignoring the privacy policies and terms of service policies of social networking services},
     volume = {23},
     issn = {1369-118X, 1468-4462},
     shorttitle = {The biggest lie on the {Internet}},
     url = {https://www.tandfonline.com/doi/full/10.1080/1369118X.2018.1486870},
     doi = {10.1080/1369118X.2018.1486870},
     language = {en},
     number = {1},
     urldate = {2022-10-14},
     journal = {Information, Communication \& Society},
     author = {Obar, Jonathan A. and Oeldorf-Hirsch, Anne},
     month = jan,
     year = {2020},
     pages = {128--147},
}

@inproceedings{Utz2019,
author = {Utz, Christine and Degeling, Martin and Fahl, Sascha and Schaub, Florian and Holz, Thorsten},
title = {(Un)Informed Consent: Studying GDPR Consent Notices in the Field},
year = {2019},
isbn = {9781450367479},
publisher = {Association for Computing Machinery},
address = {New York, NY, USA},
url = {https://doi.org/10.1145/3319535.3354212},
doi = {10.1145/3319535.3354212},
booktitle = {Proceedings of the 2019 ACM SIGSAC Conference on Computer and Communications Security},
pages = {973–990},
numpages = {18},
series = {CCS '19}
}

@article{Machuletz2020,
author = {Dominique Machuletz and Rainer Böhme},
doi = {10.2478/popets-2020-0037},
url = {https://doi.org/10.2478/popets-2020-0037},
title = {Multiple Purposes, Multiple Problems: A User Study of Consent Dialogs after GDPR},
journal = {Proceedings on Privacy Enhancing Technologies},
number = {2},
volume = {2020},
year = {2020},
pages = {481--498}
}

@inproceedings{Nouwens2020,
author = {Nouwens, Midas and Liccardi, Ilaria and Veale, Michael and Karger, David and Kagal, Lalana},
title = {Dark Patterns after the GDPR: Scraping Consent Pop-Ups and Demonstrating Their Influence},
year = {2020},
isbn = {9781450367080},
publisher = {Association for Computing Machinery},
address = {New York, NY, USA},
url = {https://doi.org/10.1145/3313831.3376321},
doi = {10.1145/3313831.3376321},
booktitle = {Proceedings of the 2020 CHI Conference on Human Factors in Computing Systems},
pages = {1–13},
numpages = {13},
series = {CHI '20}
}

@article{Grassl2021, 
title={Dark and Bright Patterns in Cookie Consent Requests}, 
volume={3}, url={https://jdsr.se/ojs/index.php/jdsr/article/view/54}, DOI={10.33621/jdsr.v3i1.54}, 
number={1}, 
journal={Journal of Digital Social Research}, 
author={Graßl, Paul and Schraffenberger, Hanna and Zuiderveen Borgesius, Frederik and Buijzen, Moniek}, 
year={2021}, 
pages={1-38} 
}

@article{OinasKukkonen2009,
  title = {Persuasive Systems Design Key Issues, Process Model, and System Features},
  author = {Harri Oinas-Kukkonen and Marja Harjumaa},
  journal = {Communications of the Association for Information Systems},
  year = {2009},
  volume = {24},
}

@misc{european_commission_regulation_2016,
  author = {{European Commission}},
  publisher = {European Commission},
  title = {Regulation ({EU}) 2016/679 of the {European} {Parliament} and of the {Council} of 27 {April} 2016 on the protection of natural persons with regard to the processing of personal data and on the free movement of such data, and repealing {Directive} 95/46/{EC} ({General} {Data} {Protection} {Regulation}) ({Text} with {EEA} relevance)},
journal = {Official Journal of the European Union L 119},
pages = {1–88},
  url = {https://eur-lex.europa.eu/eli/reg/2016/679/oj},
  year = {2016}
}
\end{document}